\documentclass[conference]{IEEEtran}
\IEEEoverridecommandlockouts

\usepackage[hidelinks]{hyperref}
\usepackage[cmex10]{amsmath}
\usepackage{amssymb,amsfonts}
\usepackage[ruled,vlined]{algorithm2e}
\usepackage{graphicx}
\graphicspath{{Figures/PDF/}{Figures/PNG/}}

\usepackage{booktabs}
\usepackage{siunitx}
\usepackage[numbers,compress]{natbib}
\usepackage{texnames}
\usepackage{bm,bbm}
\usepackage{orcidlink}
\usepackage{float}
\usepackage{multirow}
\usepackage{comment}

\begin{document}

\title{


    {Benchmarking Deep Learning and Statistical Target Detection Methods for PFM-1 Landmine Detection in UAV Hyperspectral Imagery}





}





\author{ 	\IEEEauthorblockN{
                Sagar Lekhak \orcidlink{0009-0009-7896-6167}$^1$,
                Prasanna Reddy Pulakurthi \orcidlink{0000-0003-0486-0756}$^1$,
                Ramesh Bhatta \orcidlink{0009-0005-2994-4915}$^1$,
                Emmett J. Ientilucci \orcidlink{0000-0002-3643-8245}$^1$
            } 
            \\           
	      \IEEEauthorblockA{
                \textit{$^1$Rochester Institute of Technology,}
                Chester F. Carlson Center for Imaging Science,
		      Rochester, NY 14623, USA 
            }
}



\maketitle
\begin{abstract}
In recent years, unmanned aerial vehicles (UAVs) equipped with imaging sensors and automated processing algorithms have emerged as a promising tool to accelerate large-area surveys while reducing risk to human operators. Although hyperspectral imaging (HSI) enables material discrimination using spectral signatures, standardized benchmarks for UAV-based landmine detection remain scarce. In this work, we present a systematic benchmark of four classical statistical detection algorithms, including Spectral Angle Mapper (SAM), Matched Filter (MF), Adaptive Cosine Estimator (ACE), and Constrained Energy Minimization (CEM), alongside a proposed lightweight Spectral Neural Network utilizing Parametric Mish activations for PFM-1 landmine detection. We also release pixel-level binary ground truth masks (target/background) to enable standardized, reproducible evaluation.
Evaluations were conducted on inert PFM-1 targets across multiple scene crops using a recently released VNIR hyperspectral dataset. Metrics such as receiver operating characteristic (ROC) curve, area under the curve (AUC), precision-recall (PR) curve, and average precision (AP) were used. While all methods achieve high ROC-AUC on an independent test set, the ACE method observes the highest AUC of 0.989. However, because target pixels are extremely sparse relative to background, ROC-AUC alone can be misleading; under precision-focused evaluation (PR and AP), the Spectral-NN outperforms classical detectors, achieving the highest AP. These results emphasize the need for precision-focused evaluation, scene-aware benchmarking, and learning-based spectral models for reliable UAV-based hyperspectral landmine detection. The code, dataset link, and pixel-level annotations are available at: \href{https://github.com/PrasannaPulakurthi/pfm1-hsi-benchmark}{\textit{https:/github.com/PrasannaPulakurthi/pfm1-hsi-benchmark}}.

\end{abstract}


\section{Introduction}

\begin{figure}[!t]
    \centering
    \includegraphics[width=0.9\linewidth]{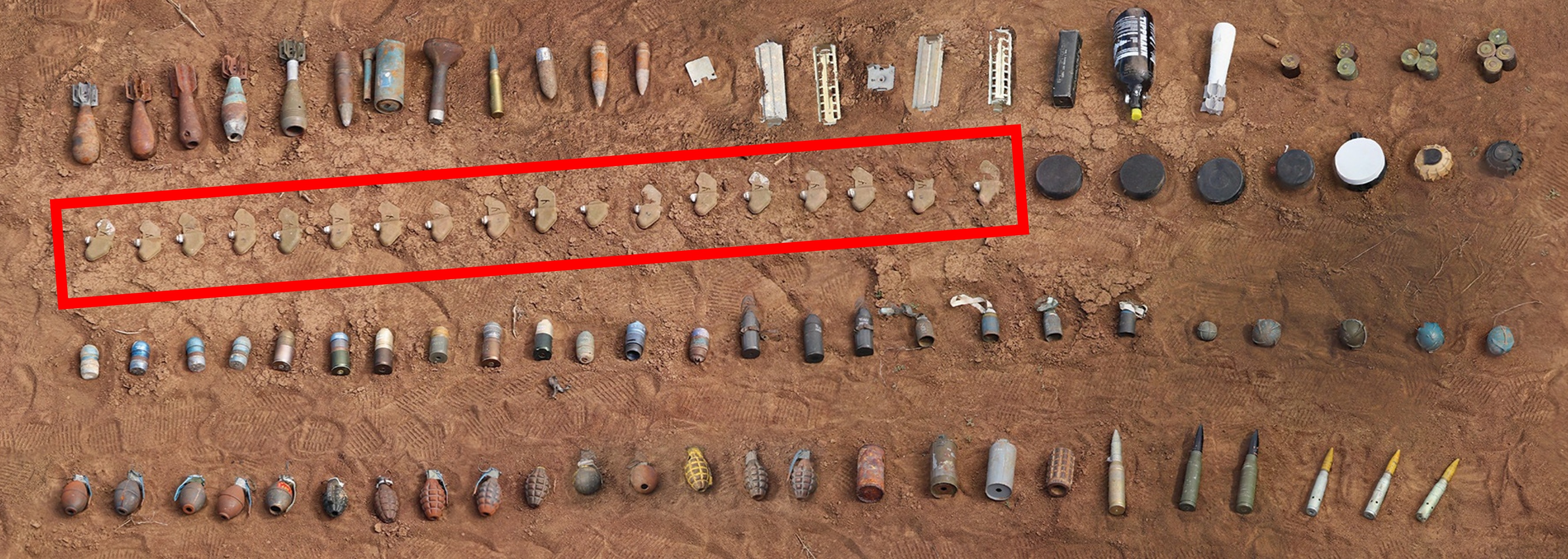}
    \vspace{-0.1in}
    \caption{Overview of PFM-1 mines (highlighted in red) and other target types (Image source \cite{baur2023accessible}).}
    \label{fig:intro} 
    \vspace{-0.25in}
\end{figure}

Surface-laid landmines and unexploded ordnance (UXO) remain a serious humanitarian and security challenge. To address this, various deep learning approaches have been proposed using RGB imagery \cite{vivoli, stankevich2024workflow, saprykin2024optical, baur2021drones}. However, as illustrated in Fig~\ref{fig:intro}, targets like the PFM-1 landmine are specifically designed to be visually indistinguishable from natural backgrounds. This inherent camouflage means that the limited spectral content of standard RGB sensors often hampers the effective discrimination of landmines from their surroundings~\cite{WANG2024103645}. In addition, Lekhak \emph{et al.} \cite{lekhak2025uncertainty} analyzed uncertainty and adversarial sensitivity in RGB-based landmine classification, and other studies have reported performance degradation under poor illumination, occlusion, and complex outdoor conditions \cite{rgb_failure_poor_light}.

To overcome the limitations of RGB imagery, recent research has progressed toward utilizing hyperspectral imaging (HSI). Unlike RGB sensors, HSI captures hundreds of narrow, contiguous spectral bands, enabling the discrimination of materials based on subtle spectral signatures that are often indistinguishable in the visible spectrum. While foundational work surveyed statistical target and anomaly detection methods for landmine remediation \cite{MAKKI201740}, the majority of prior studies focused on laboratory setups, ground-based sensors, or manned airborne platforms, limiting their applicability to rapid-response UAV deployments.

UAV-mounted hyperspectral systems offer significant operational potential for surface landmine detection, effectively reducing the time, cost, and physical risk associated with humanitarian demining \cite{HyperspectralNTS}. While recent studies have demonstrated the feasibility of detecting surficial explosive ordnance \cite{Tuohy} and assessed the data quality of drone-mounted hyperspectral sensors \cite{data_quality_assessment}, these efforts have primarily emphasized system deployment and qualitative results. Consequently, a critical gap remains: the lack of standardized benchmarks and pixel-level ground truth to objectively compare algorithmic performance in these unique aerial scenarios.

This need for benchmarking is further complicated by the inherent nature of the landmine detection task. While modern transformer and attention-based architectures \cite{Chatterjee_2024, Chatterjee_2025} have significantly advanced hyperspectral image classification, they typically rely on abundant labeled data that are rarely met in environments where targets are extremely sparse. In contrast, classical target detection methods such as SAM \cite{sam}, MF \cite{mf}, ACE \cite{ace_original, ace_hsi_popular}, and CEM \cite{cem_original_concept, cem_hsi_popular} remain widely used due to interpretability and minimal training needs. However, despite their widespread use, these traditional detectors have not been systematically benchmarked on modern UAV-based hyperspectral datasets, particularly when facing the complex spectral and spatial challenges posed by PFM-1 targets.

This paper addresses these gaps using a recently released UAV-based VNIR hyperspectral benchmark dataset \cite{lekhak2025uav}, containing over 140 inert landmine-type targets in realistic outdoor settings. 
In this work, we focus specifically on surface PFM-1 landmines targets in the dataset, shown in Fig. \ref{fig:intro} as a challenging case study due to their small size, plastic composition, and spectral similarity to natural backgrounds.

The main contributions of this work are as follows:

\begin{enumerate}
    \item \textbf{Systematic Benchmark:} We systematically benchmarked classical target detection algorithms for UAV-based PFM-1 landmine detection, with insights from ROC-AUC and Average Precision (AP) showing the importance of precision-focused evaluation. 
    \item \textbf{Proposed Spectral-NN:} We demonstrated that a simple proposed spectral neural network outperformed statistical target detection algorithms for the dataset used.
    \item \textbf{Dataset Release:} To support reproducible research, we released pixel-level binary ground truth masks for PFM-1 targets. This release includes curated, cropped sections of the dataset containing only mines or PFM-1 targets, covering both surface-visible and partially buried cases.
\end{enumerate}

By establishing a benchmark and highlighting classical and deep learning approaches, this work motivates further research toward operational UAV hyperspectral landmine detection.



\section{Methodology}

\subsection{Preprocessing and Ground Truth Annotation}

\begin{figure}[!t]
    \centering
    \includegraphics[width=0.95\linewidth]{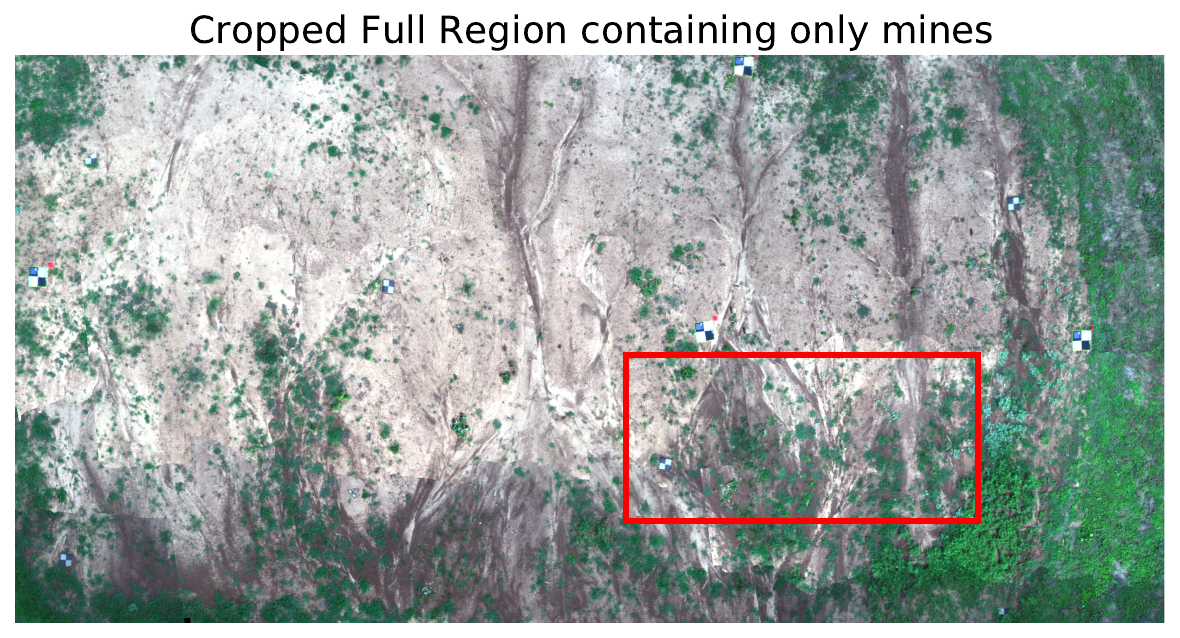}
    \includegraphics[width=0.95\linewidth]{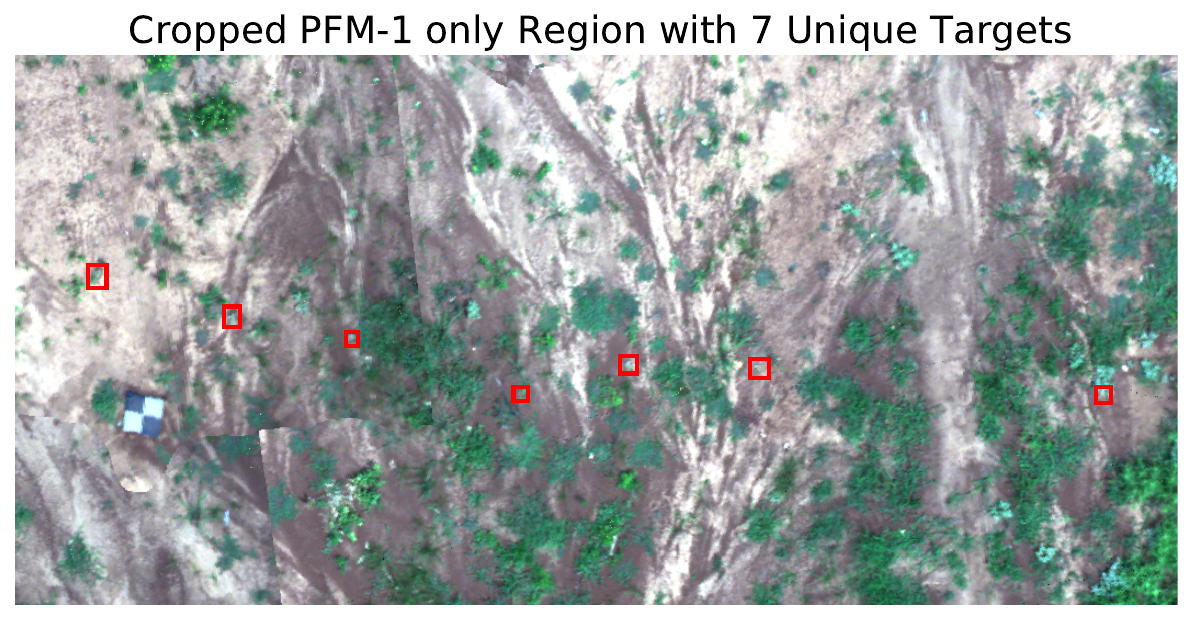}
    \includegraphics[width=0.95\linewidth]{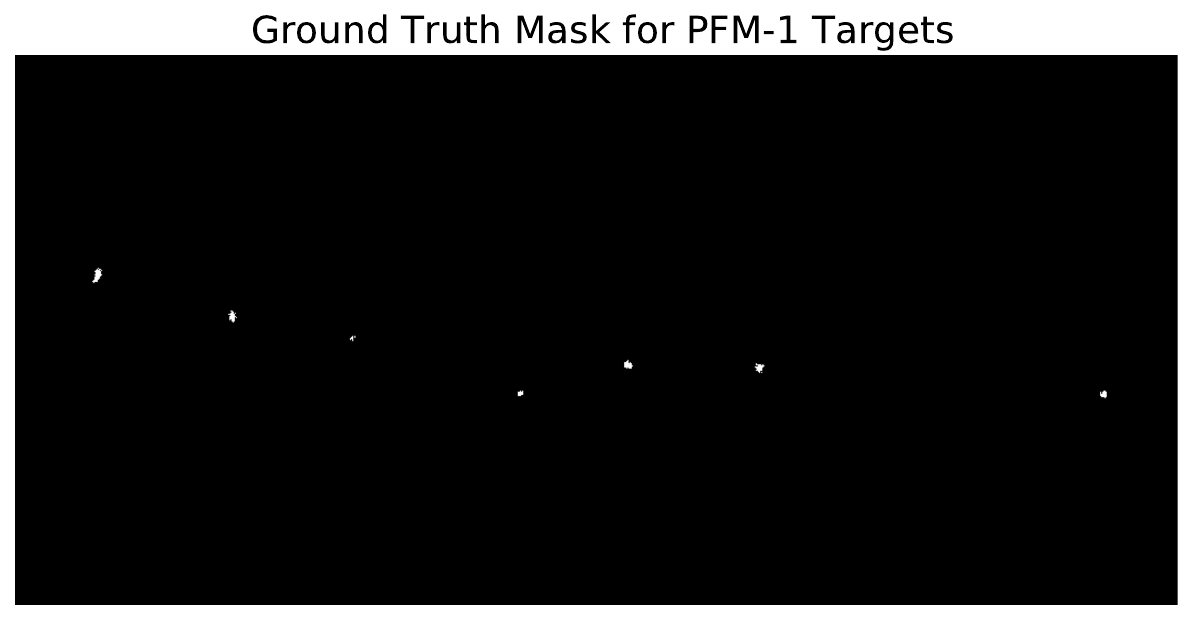}
    \vspace{-0.1in}
    \caption{Cropped regions used in this study from the original hyperspectral dataset \cite{lekhak2025uav}: (top) \textit{Full Region}, containing all mine locations; (middle) \textit{PFM-1 Region}, showing only PFM-1 targets; (bottom) Ground truth binary mask of the \textit{PFM-1 Region}, with training set containing the first five PFM-1 targets (left) and the remaining two targets used for testing (right). GSD $\approx$ 1.29 cm.}
    \label{fig:rgb_image_and_mask}
    \vspace{-0.1in}
\end{figure}

The UAV-based hyperspectral dataset presented in \cite{lekhak2025uav} contains a variety of inert landmine and UXO objects imaged using a VNIR hyperspectral sensor. For this study, we focus exclusively on PFM-1 landmines to enable a controlled evaluation of target detection performance. Starting from the original hyperspectral cube (3123 rows $\times$ 6631 samples $\times$ 272 spectral bands) provided by the authors, the data were cropped spatially in three stages, retaining all the bands, as shown in Fig.~\ref{fig:rgb_image_and_mask}. First, we retained only mine-containing regions (1705 $\times$ 3461), referred to as the \emph{Full Region}. Next, we isolated only PFM-1 targets, referred to as the \emph{PFM-1 Region}, resulting in dataset of spatial dimension 500 $\times$ 1060. Finally, this \emph{PFM-1 Region} was further divided into independent training and test regions. The training region spans from the left of the ground truth mask shown at the bottom figure of Fig.~\ref{fig:rgb_image_and_mask} with dimensions 500 $\times$ 610, and contains the first five PFM-1 targets, while the \textit{Test Region} spans across the remaining target regions with dimensions 500 $\times$ 450. This ensures no data leakage for testing of the proposed neural network. 

%

For an unbiased comparison between all classical algorithms and the proposed neural network, all methods were evaluated in the \textit{Test Region}. In both \textit{Full Region} and \textit{PFM-1 Region}, the classical algorithms were assessed for their behavior in large-scale scenarios involving millions of pixels.         





Seven locations containing partially or fully visible PFM-1 mines were manually annotated in ENVI\textsuperscript{\textregistered} \cite{ENVI} \footnote{ENVI\textsuperscript{\textregistered} is a registered trademark of NV5 Global, Inc.} to generate a binary ground truth mask (1 for target, 0 for background). This resulted in 248 target pixels, using georeferenced hyperspectral and high-resolution RGB imagery for improved accuracy. 


\subsection{Statistical Target Detection Algorithms}

Classical statistical target detection algorithms exploit spectral similarity and background statistics to detect a known target signature in hyperspectral imagery. Four widely used detectors, SAM~\cite{sam}, MF~\cite{mf}, ACE~\cite{ace_original, ace_hsi_popular}, and CEM~\cite{cem_original_concept, cem_hsi_popular}, are evaluated using the ground spectral signature of PFM-1 provided by \cite{lekhak2025uav}. In our implementation of SAM, a negative sign was applied to the scores so that higher values correspond to stronger target responses. For consistency across MF, ACE, and CEM, the data were mean-centered, and the resulting detection scores were normalized to the range $[0,1]$ for visualization and comparison.

\subsection{Deep Learning: Spectral Neural Network}

To complement statistical detectors, a deep learning model was proposed for supervised target-versus-background classification using only spectral information. Each pixel, represented as a 272-dimensional vector, is treated independently. The fully connected network features two hidden layers (128 and 64 neurons) with Parametric Mish (PMish) \cite{pmish} activations and a single-neuron sigmoid output layer for target probability. PMish was selected due to its smoother gradients and better convergence \cite{pmish}. Class imbalance was addressed via a positive class weight in a weighted binary cross-entropy loss~\cite{ho2019real}. Training was conducted for 50 epochs using the Adam optimizer \cite{kingma2014adam} with a learning rate of 0.0002.
%
%
%
Formally, given an input spectral vector $\mathbf{x} \in \mathbb{R}^{272}$, the network computes:
\begin{align}
&\mathbf{h}_1 = \mathrm{PMish}(\mathbf{W}_1 \mathbf{x} + \mathbf{b}_1), \\
&\mathbf{h}_2 = \mathrm{PMish}(\mathbf{W}_2 \mathbf{h}_1 + \mathbf{b}_2), \\
&D_{\text{DL}}(\mathbf{x}) = \sigma(\mathbf{W}_3 \mathbf{h}_2 + \mathbf{b}_3),
\end{align}

where $\sigma(\cdot)$ denotes the sigmoid function. The output $D_{\text{DL}}(\mathbf{x}) \in [0,1]$ represents the probability of the pixel belonging to the PFM-1 target class.

\subsection{Evaluation Metrics}

To evaluate detection performance, we employed two complementary threshold-independent metrics: Area Under the Curve (AUC), derived from the Receiver Operating Characteristic (ROC) curve, and Average Precision (AP), derived from the precision--recall (PR) curve. ROC analysis provides a threshold-independent measure of separability between target and background pixels; however, it can be overly optimistic in scenarios where background pixels vastly outnumber target pixels. In contrast, AP emphasizes the quality of positive detections by integrating precision across all recall levels, making it more sensitive to false alarms in target-sparse environments. Reporting both metrics, therefore, offers a comprehensive assessment: ROC-AUC reflects global discrimination capability, while AP captures practical detection reliability in operational UAV-based landmine detection.

\section{Results and Discussion}




\begin{figure*}[!t]
    \centering
    \includegraphics[width=0.89\linewidth]{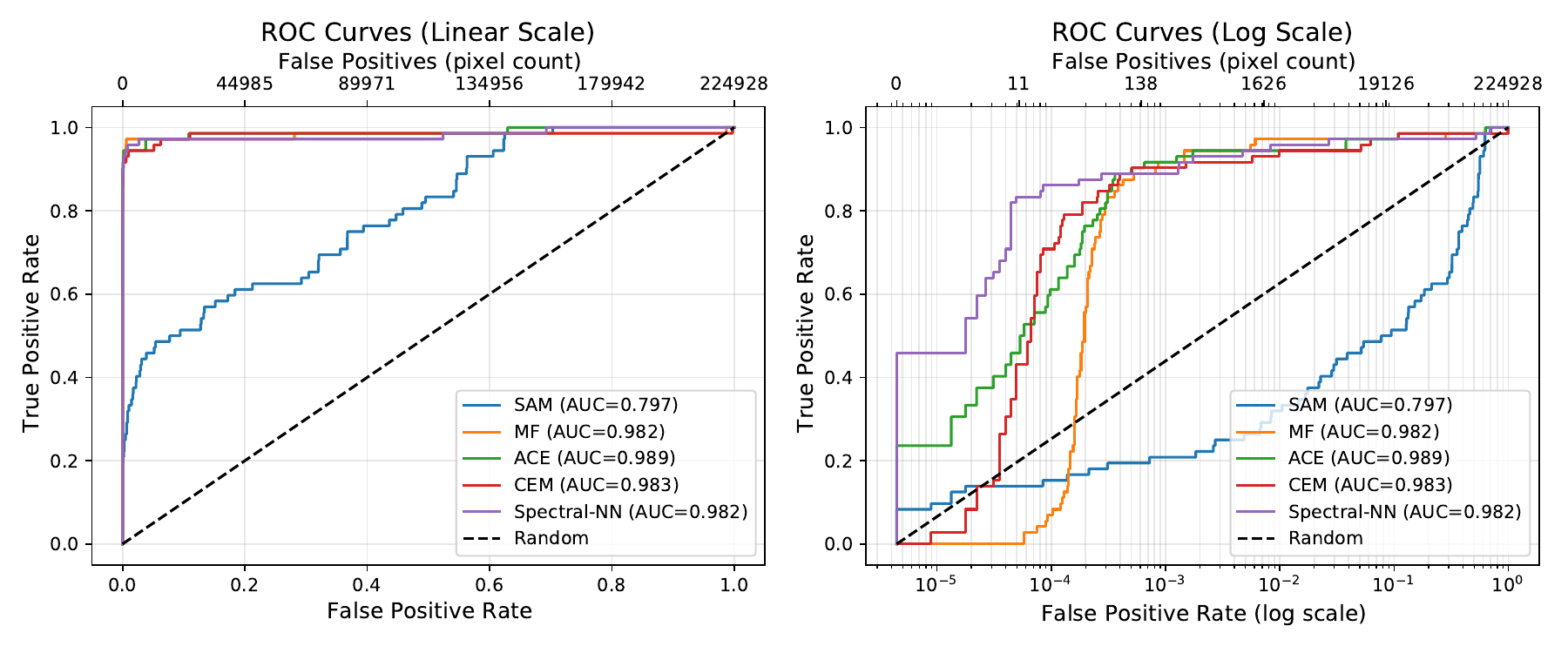}
    \vspace{-0.2in}
    \caption{ROC curves and corresponding AUCs for all algorithms in the \textit{Test Region}: a) linear scale and b) logarithmic scale.}
    \label{fig:roc_results}
    \vspace{-0.1in}
\end{figure*}

\begin{figure*}[!t]
    \centering
    \includegraphics[width=0.309\linewidth]{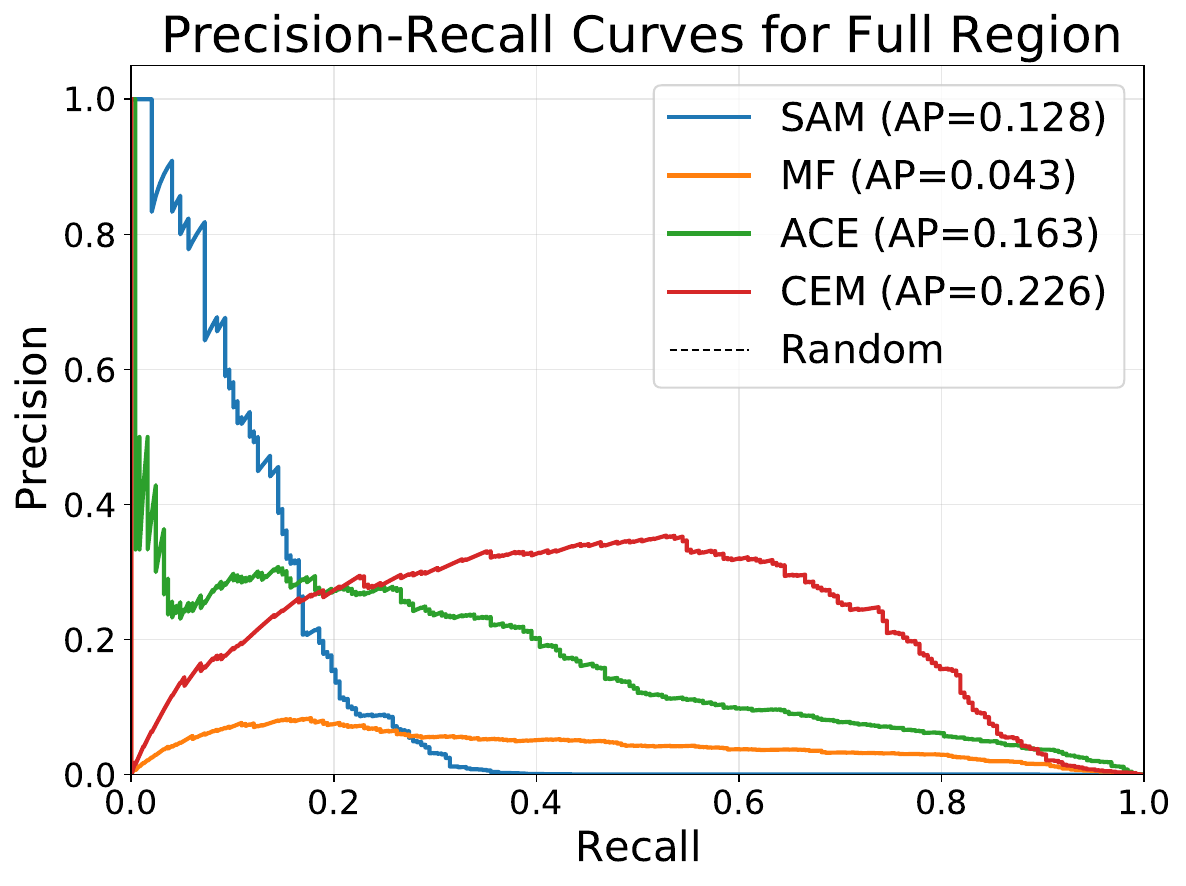}
    \includegraphics[width=0.339\linewidth]{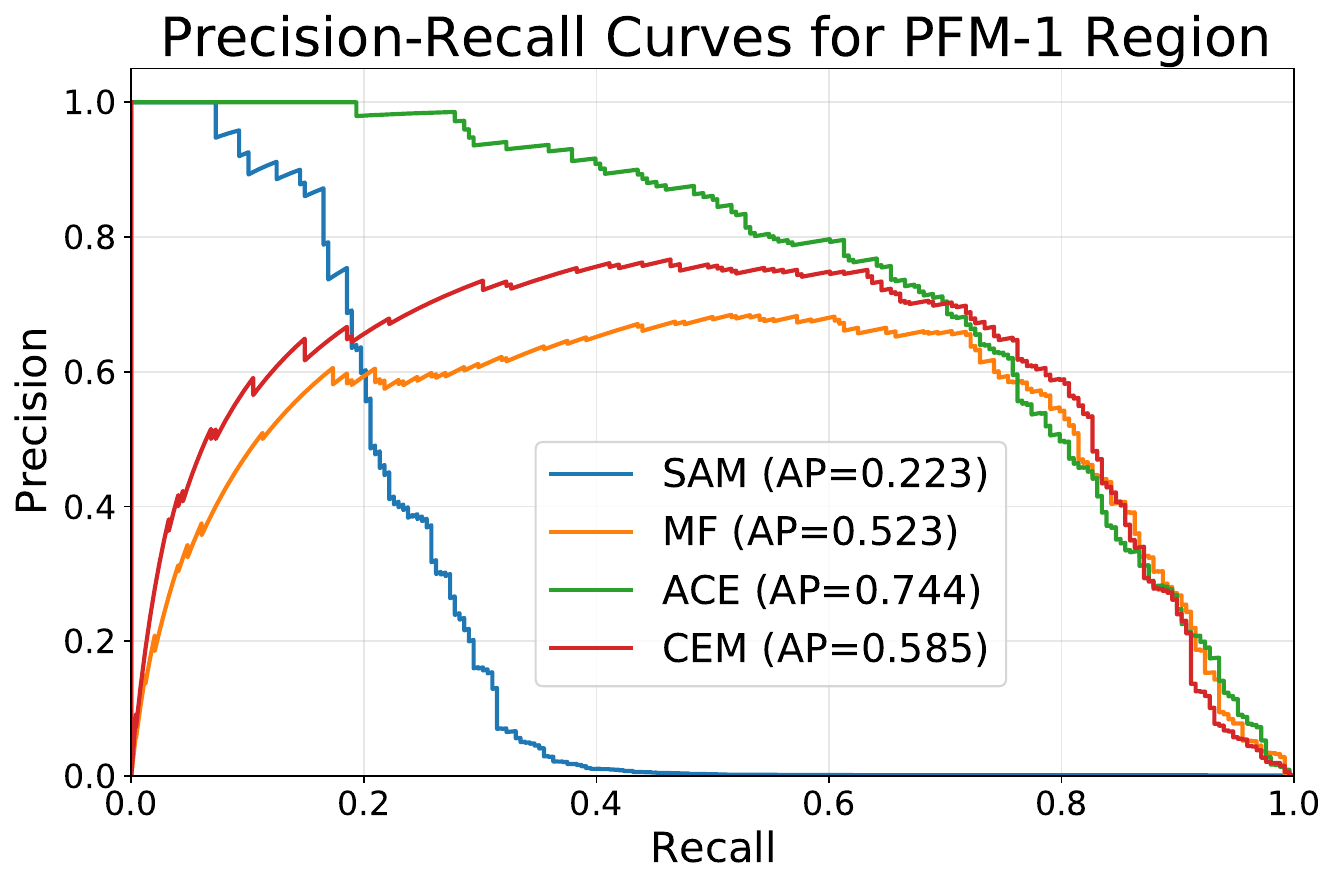}
    \includegraphics[width=0.339\linewidth]{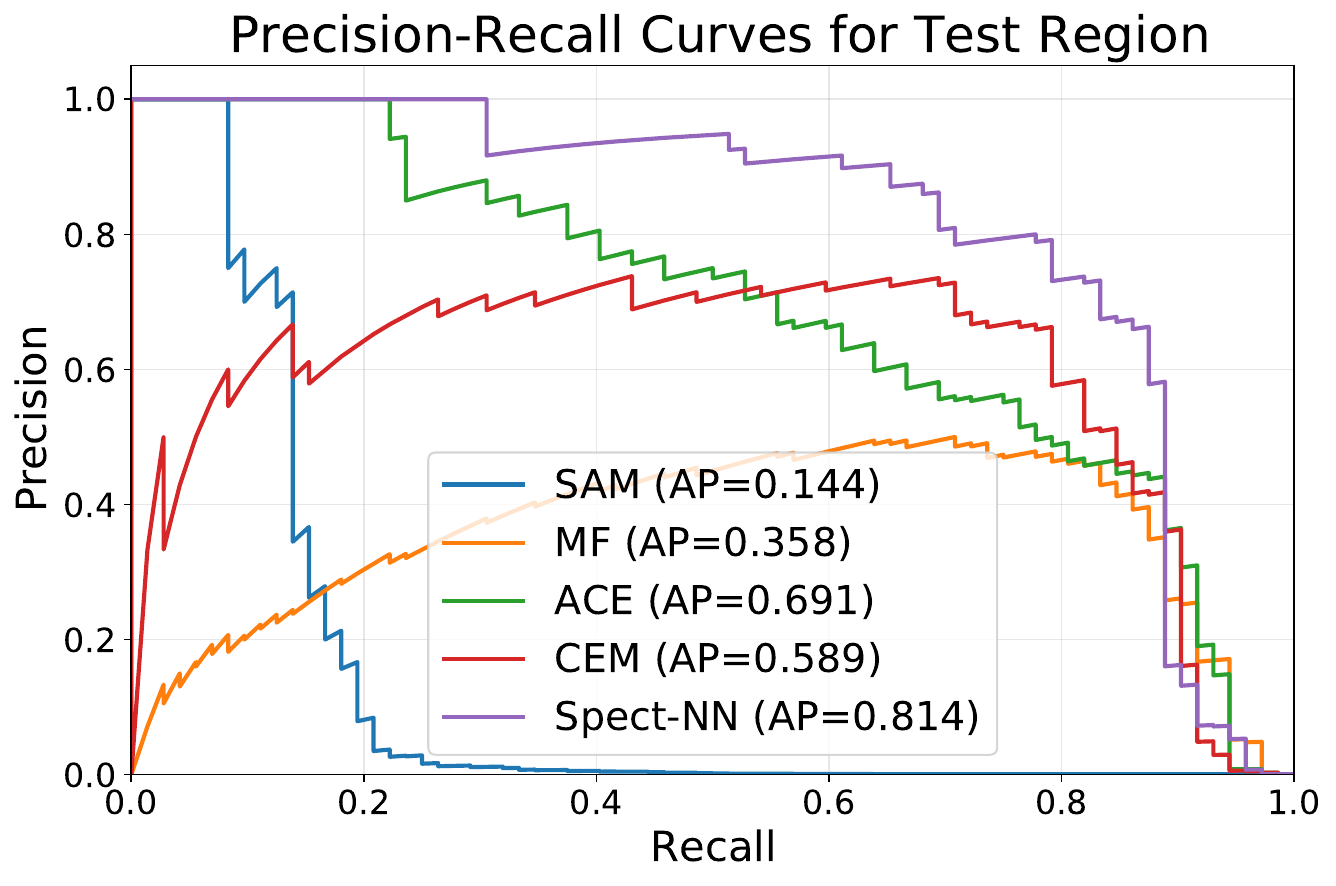}
    \vspace{-0.25in}
    \caption{Precision-recall curve and corresponding APs for a) \textit{Full Region} (left), b) \textit{PFM-1 Region} (middle), and \textit{Test Region} (right).} 
    \label{fig:pr_curve}
    \vspace{-0.1in}
\end{figure*}

\begin{figure}[!t]
    \centering
    \includegraphics[width=0.95\linewidth]{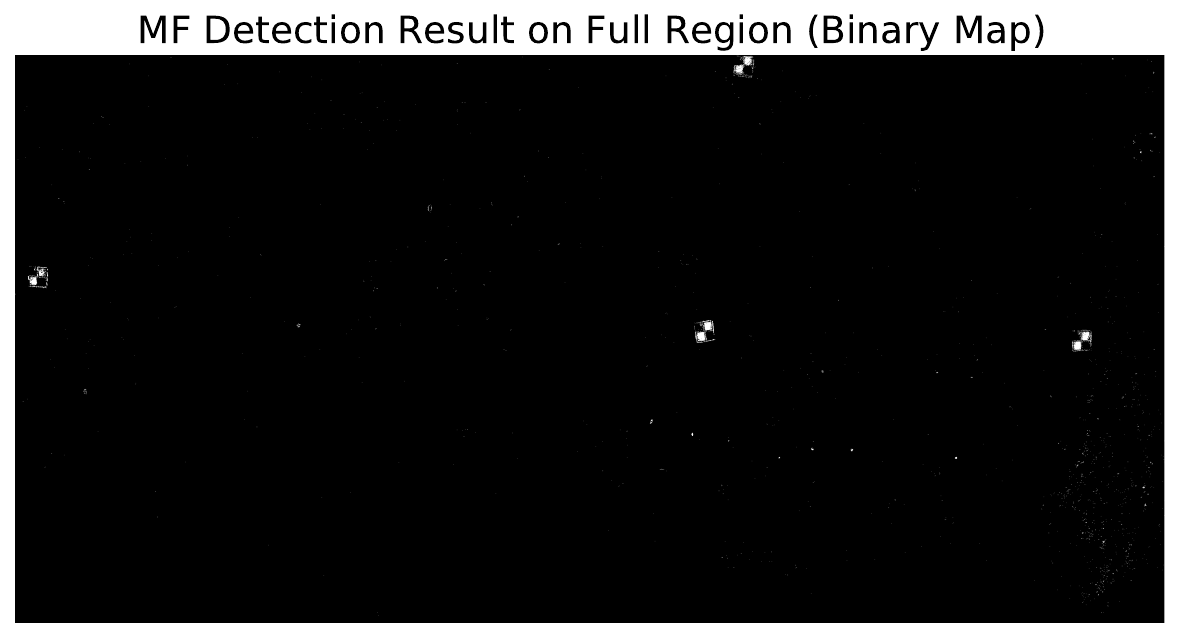}
    \vspace{-0.1in}
    \caption{Aeropoints in the scene appearing as the biggest false alarms for PFM-1 mines even at higher threshold values for detection scores in classical statistical methods in \textit{Full Region}. Presented here is an example case for a representative thresholds (0.77 -- 0.85) in MF detection scores.}
    \label{fig:mf_detection_plot}
    \vspace{-0.2in}
\end{figure}

\begin{table}[!t]
\centering
\caption{Performance Comparison of Detection Models}
\label{tab:model_results}
\renewcommand{\arraystretch}{1.2}
\resizebox{0.9\columnwidth}{!}{  
\begin{tabular}{l|cc|cc|cc}
\hline

\hline
\multicolumn{1}{c|}{\multirow{2}{*}{\textbf{Model}}} & \multicolumn{2}{c|}{\textbf{Full Region}} & \multicolumn{2}{c|}{\textbf{PFM-1 Region}} & \multicolumn{2}{c}{\textbf{Test Region}} \\ \cline{2-7} 
\multicolumn{1}{c|}{} & AP & AUC & AP & AUC & AP & AUC \\ 
\hline

\hline
SAM \cite{sam} & 0.128 & 0.718 & 0.223 & 0.741 & 0.144 & 0.797 \\
MF \cite{mf} & 0.043 & 0.994 & 0.523 & 0.994 & 0.358 & 0.982 \\
ACE \cite{ace_hsi_popular, ace_original} & 0.163 & \textbf{0.998} & \textbf{0.744} & \textbf{0.997} & 0.691 & \textbf{0.989} \\
CEM \cite{cem_original_concept, cem_hsi_popular} & \textbf{0.226} & 0.995 & 0.585 & 0.995 & 0.589 & 0.983 \\ 
Spectral-NN & -- & -- & -- & -- & \textbf{0.814} & 0.982 \\ 
\hline

\hline
\end{tabular}}
\vspace{-0.1in}
\end{table}

Table~\ref{tab:model_results} summarizes the performance of four classical statistical target detection algorithms and the proposed multilayer perceptron (Spectral-NN) across three evaluation regions: the \textit{Full Region}, the cropped \textit{PFM-1 Region}, and the independent \textit{Test Region} used for unbiased assessment. 

\subsection{ROC-Based Analysis}

ROC analysis, presented in Fig. \ref{fig:roc_results} for the independent \textit{Test Region}, shows that ACE, MF, CEM, and Spectral-NN achieve similarly high AUC values on the linear scale (approximately 0.98). However, examination of the log-scale ROC curves reveals substantial differences in detector behavior within the operationally critical low false-positive regime. ACE consistently attains higher true positive rates at very low false-positive rates, indicating superior separation between target and background score distributions and increased robustness to background variability. In contrast, MF and CEM exhibit delayed detection onset at low thresholds, suggesting greater overlap between target and background responses despite strong global discrimination. Spectral-NN demonstrates competitive early detection performance but does not consistently outperform ACE across the entire low-FPR range, while SAM performs near a lower-bound baseline. These results highlight that aggregate AUC alone is insufficient for evaluating rare-target detection performance and that low-FPR behavior is a key discriminator for landmine detection applications.

\subsection{Precision--Recall Analysis}

Precision--recall analysis, presented in Fig. \ref{fig:pr_curve} provides complementary insight under severe class imbalance and reveals detector behavior not apparent from ROC curves alone. For the \textit{Full Region}, all methods exhibit low Average Precision (AP), reflecting extreme target sparsity and high background diversity. In this setting, CEM achieves the highest AP, followed by ACE and SAM, while MF performs poorly across most recall levels. When evaluating on the \textit{PFM-1 Region}, background variability is reduced, and all detectors show substantial AP improvement. In this case, ACE achieves the highest AP, followed by CEM and MF, indicating more stable precision across a wider recall range. SAM exhibits high precision only at very low recall levels and degrades rapidly, confirming its limited robustness to background clutter. These results demonstrate that PR metrics are crucial for assessing detector reliability in highly imbalanced landmine detection scenarios.

\subsection{Impact of Scene Composition and Cropping}

Table~\ref{tab:model_results} further illustrates the strong dependence of detection performance on scene composition. In the \textit{Full Region}, numerous false positives originate from non-target objects and aeropoints (ground control points) as illustrated in Fig.~\ref{fig:mf_detection_plot}. Upon inspection, the spectral signatures of these false alarms are not similar to PFM-1's target. The reason behind their appearance will be an area of future investigation. These false positives lead to reduced average precision (AP) despite high AUC values in \textit{Full Region}. Cropping the scene to the \textit{PFM-1 Region} removes many of these confounding structures, resulting in a significant increase in AP across all methods. On the independent \textit{Test Region}, ACE achieves the best balance between AP (0.691) and AUC (0.989), demonstrating strong generalization to unseen regions. CEM and MF exhibit comparable AUC values but reduced AP, indicating sensitivity to local background statistics. SAM consistently underperforms across all evaluation settings.

\subsection{Comparison with Deep Learning}

The Spectral-NN achieves the strongest precision--recall balance on the independent \textit{Test Region}, while maintaining competitive ROC performance, as illustrated in Fig. \ref{fig:roc_results} and Fig. \ref{fig:pr_curve}. This result demonstrates that even a lightweight spectral neural network can surpass classical statistical detectors when trained on carefully curated hyperspectral signatures. These findings suggest that learning-based spectral models offer a promising direction for detecting challenging surface landmines such as PFM-1 under realistic UAV operating conditions.

Overall, the results confirm that (i) ROC-AUC alone is insufficient for assessing rare-target detection performance, (ii) precision-based metrics are strongly influenced by background variability and scene composition, and (iii) ACE provides the most robust classical performance, while the proposed Spectral-NN demonstrates superior generalization and practical detection reliability.

\section{Conclusion and Future Directions}
This work presented a systematic benchmark of classical statistical target detection methods and a lightweight spectral neural network for PFM-1 landmine detection at different cropped regions in UAV-based hyperspectral imagery. The results demonstrate that ROC-AUC metric alone might not be sufficient in highly imbalanced, rare-target scenarios and that precision–recall metrics are essential for assessing detection reliability. Among classical approaches, ACE provided the most robust performance, while the spectral neural network achieved the strongest precision–recall balance. 
The study also demonstrates the feasibility of UAV-based hyperspectral systems for surface landmine detection compared to RGB-based methods, which offer limited spectral information. 

Future work will investigate uncertainty-aware and SOTA spatial–spectral models to improve robustness and operational applicability. In addition, the availability of a co-registered drone-based metal detector dataset over the same test site in \cite{lekhak_rs} enables exploration of multi-sensor data fusion for detecting both surface and shallowly buried PFM-1 landmines.

\newpage
\small
\bibliographystyle{IEEEtranN}
\bibliography{references}

@inproceedings{stankevich2024workflow,
  title={Workflow for Landmine Detection in Aerial Imagery with YOLO-based Deep Learning},
  author={Stankevich, SA and Golubov, SI and Dugin, SS and Saprykin, IY},
  booktitle={2024 IEEE 7th International Conference on Actual Problems of Unmanned Aerial Vehicles Development (APUAVD)},
  pages={82--86},
  year={2024},
  organization={IEEE}
}

@article{saprykin2024optical,
  title={OPTICAL DEEP LEARNING LANDMINE DETECTION BASED ON LIMITED DATASET OF AERIAL IMAGERY.},
  author={Saprykin, Ievgen},
  journal={Science-based technologies},
  volume={62},
  number={2},
  year={2024}
}

@article{lekhak2025uav,
  title={A UAV-Based VNIR Hyperspectral Benchmark Dataset for Landmine and UXO Detection},
  author={Lekhak, Sagar and Ientilucci, Emmett J and Baur, Jasper and Ghosh, Susmita},
  journal={arXiv preprint arXiv:2510.02700},
  year={2025}
}

@article{sam,
title = {The spectral image processing system (SIPS)—interactive visualization and analysis of imaging spectrometer data},
journal = {Remote Sensing of Environment},
volume = {44},
number = {2},
pages = {145-163},
year = {1993},
note = {Airbone Imaging Spectrometry},
issn = {0034-4257},
doi = {https://doi.org/10.1016/0034-4257(93)90013-N},
url = {https://www.sciencedirect.com/science/article/pii/003442579390013N},
author = {F.A. Kruse and A.B. Lefkoff and J.W. Boardman and K.B. Heidebrecht and A.T. Shapiro and P.J. Barloon and A.F.H. Goetz}
}

@ARTICLE{mf,
  author={Manolakis, D. and Shaw, G.},
  journal={IEEE Signal Processing Magazine}, 
  title={Detection algorithms for hyperspectral imaging applications}, 
  year={2002},
  volume={19},
  number={1},
  pages={29-43},
  doi={10.1109/79.974724}}

@ARTICLE{ace_original,
  author={Kraut, S. and Scharf, L.L.},
  journal={IEEE Transactions on Signal Processing}, 
  title={The CFAR adaptive subspace detector is a scale-invariant GLRT}, 
  year={1999},
  volume={47},
  number={9},
  pages={2538-2541},
  doi={10.1109/78.782198}}

@ARTICLE{ace_hsi_popular,
  author={Kraut, S. and Scharf, L.L. and Butler, R.W.},
  journal={IEEE Transactions on Signal Processing}, 
  title={The adaptive coherence estimator: a uniformly most-powerful-invariant adaptive detection statistic}, 
  year={2005},
  volume={53},
  number={2},
  pages={427-438},
  doi={10.1109/TSP.2004.840823}}

@phdthesis{cem_original_concept,
  title={Detection and Classification of Subpixel Spectral Signatures in Hyperspectral Image Sequences},
  author={Harsanyi, Joseph C.},
  year={1993},
  school={University of Maryland, Baltimore County},
  address={Baltimore, MD}
}

@article{cem_hsi_popular,
title = {Mapping the distribution of mine tailings in the Coeur d'Alene River Valley, Idaho, through the use of a constrained energy minimization technique},
journal = {Remote Sensing of Environment},
volume = {59},
number = {1},
pages = {64-76},
year = {1997},
issn = {0034-4257},
doi = {https://doi.org/10.1016/S0034-4257(96)00080-6},
url = {https://www.sciencedirect.com/science/article/pii/S0034425796000806},
author = {William H. Farrand and Joseph C. Harsanyi}
}

@ARTICLE{pmish,
  author={Pulakurthi, Prasanna Reddy and Mozaffari, Mahsa and Dianat, Sohail A. and Heard, Jamison and Rao, Raghuveer M. and Rabbani, Majid},
  journal={IEEE Access}, 
  title={Enhancing GANs With MMD Neural Architecture Search, PMish Activation Function, and Adaptive Rank Decomposition}, 
  year={2024},
  volume={12},
  number={},
  pages={174222-174244},
  doi={10.1109/ACCESS.2024.3485557},
  url={https://ieeexplore.ieee.org/document/10732016}}

@article{Tuohy,
author = {Tuohy, Madison and Baur, Jasper and Steinberg, Gabriel and Pirro, Jalissa and Mitchell, Taylor and Nikulin, Alex and Frucci, John and Smet, Timothy},
year = {2023},
month = {02},
pages = {98-102},
title = {Utilizing UAV-based hyperspectral imaging to detect surficial explosive ordnance},
volume = {42},
journal = {The Leading Edge},
doi = {10.1190/tle42020098.1}
}

@article{MAKKI201740,
title = {A survey of landmine detection using hyperspectral imaging},
journal = {ISPRS Journal of Photogrammetry and Remote Sensing},
volume = {124},
pages = {40-53},
year = {2017},
issn = {0924-2716},
doi = {https://doi.org/10.1016/j.isprsjprs.2016.12.009},
url = {https://www.sciencedirect.com/science/article/pii/S0924271616306451},
author = {Ihab Makki and Rafic Younes and Clovis Francis and Tiziano Bianchi and Massimo Zucchetti}
}

@inproceedings{data_quality_assessment,
author = {Randall A. Pietersen and Junhyung Park and Jonathan W. Reasoner and Brian M. Robinson and Robert A. Diltz and Herbert H. Einstein},
title = {{Data quality assessment of drone-mounted hyperspectral imaging system for unexploded ordnance (UXO) detection}},
volume = {13031},
booktitle = {Algorithms, Technologies, and Applications for Multispectral and Hyperspectral Imaging XXX},
editor = {Miguel Velez-Reyes and David W. Messinger},
organization = {International Society for Optics and Photonics},
publisher = {SPIE},
pages = {1303108},
year = {2024},
doi = {10.1117/12.3013143},
URL = {https://doi.org/10.1117/12.3013143}
}

@Article{lekhak_rs,
AUTHOR = {Lekhak, Sagar and Ientilucci, Emmett J. and Brinkley, Anthony Wayne},
TITLE = {Viability of Substituting Handheld Metal Detectors with an Airborne Metal Detection System for Landmine and Unexploded Ordnance Detection},
JOURNAL = {Remote Sensing},
VOLUME = {16},
YEAR = {2024},
NUMBER = {24},
ARTICLE-NUMBER = {4732},
URL = {https://www.mdpi.com/2072-4292/16/24/4732},
ISSN = {2072-4292},
DOI = {10.3390/rs16244732}
}

@InProceedings{Chatterjee_2025,
    author    = {Chatterjee, Abhiroop and Ghosh, Susmita and Ghosh, Ashish},
    title     = {Context-Aware Masking and Learnable Diffusion-Guided Patch Refinement in Transformers via Sparse Supervision for Hyperspectral Image Classification},
    booktitle = {Proceedings of the IEEE/CVF International Conference on Computer Vision (ICCV) Workshops},
    month     = {October},
    year      = {2025},
    pages     = {2906-2915}
}

@INPROCEEDINGS{Chatterjee_2024,
  author={Chatterjee, Abhiroop and Ghosh, Susmita and Ghosh, Ashish and Ientilucci, Emmett},
  booktitle={IGARSS 2024 - 2024 IEEE International Geoscience and Remote Sensing Symposium}, 
  title={Urbanscape-Net: A Spatial and Self-Attention Guided Deep Neural Network with Multi Scale Feature Extraction for Urban Land-Use Classification}, 
  year={2024},
  volume={},
  number={},
  pages={4884-4889},
  doi={10.1109/IGARSS53475.2024.10640965}}

@INPROCEEDINGS{lekhak2025uncertainty,
  author={Lekhak, Sagar and Ientilucci, Emmett J. and Dera, Dimah and Ghosh, Susmita},
  booktitle={IGARSS 2025 - 2025 IEEE International Geoscience and Remote Sensing Symposium}, 
  title={Uncertainty Quantification in Surface Landmines and UXO Classification using MC Dropout}, 
  year={2025},
  volume={},
  number={},
  pages={1177-1181},
  doi={10.1109/IGARSS55030.2025.11243138}}

@article{baur2021drones,
  author = {Baur, Gabriel and Steinberg, Jasper and Nikulin, Alex and Chiu, Kevin and de Smet, Timothy},
  title = {How to implement drones and machine learning to reduce time, costs, and dangers associated with landmine detection},
  journal = {The Journal of Conventional Weapons Destruction},
  volume = {25},
  number = {1},
  year = {2021},
  note = {Online. Available: \url{https://commons.lib.jmu.edu/cisr-journal/vol25/iss1/29}},
}

@Article{vivoli,
AUTHOR = {Vivoli, Emanuele and Bertini, Marco and Capineri, Lorenzo},
TITLE = {Deep Learning-Based Real-Time Detection of Surface Landmines Using Optical Imaging},
JOURNAL = {Remote Sensing},
VOLUME = {16},
YEAR = {2024},
NUMBER = {4},
ARTICLE-NUMBER = {677},
URL = {https://www.mdpi.com/2072-4292/16/4/677},
ISSN = {2072-4292},
DOI = {10.3390/rs16040677}
}

@article{baur2023accessible,
  title={An accessible seeded field for humanitarian mine action research},
  author={Baur, Jasper and Steinberg, Gabriel and Frucci, John and Brinkley, Anthony},
  journal={The Journal of Conventional Weapons Destruction},
  volume={27},
  number={3},
  pages={2},
  year={2023}
}

@ARTICLE{rgb_failure_poor_light,
  author={Wang, Qingwang and Sun, Yuxuan and Chi, Yongke and Shen, Tao},
  journal={IEEE Journal of Selected Topics in Applied Earth Observations and Remote Sensing}, 
  title={RGB-T Object Detection With Failure Scenarios}, 
  year={2025},
  volume={18},
  number={},
  pages={3000-3010},
  doi={10.1109/JSTARS.2024.3523408}}

@article{WANG2024103645,
title = {Deep learning-based spectral reconstruction in camouflaged target detection},
journal = {International Journal of Applied Earth Observation and Geoinformation},
volume = {126},
pages = {103645},
year = {2024},
issn = {1569-8432},
doi = {https://doi.org/10.1016/j.jag.2023.103645},
url = {https://www.sciencedirect.com/science/article/pii/S1569843223004697},
author = {Shu Wang and Yixuan Xu and Dawei Zeng and Feng Huang and Lingyu Liang}
}

@article{kingma2014adam,
  title={Adam: A method for stochastic optimization},
  author={Kingma, Diederik P},
  journal={arXiv preprint arXiv:1412.6980},
  year={2014}
}

@article{ho2019real,
  title={The real-world-weight cross-entropy loss function: Modeling the costs of mislabeling},
  author={Ho, Yaoshiang and Wookey, Samuel},
  journal={IEEE access},
  volume={8},
  pages={4806--4813},
  year={2019},
  publisher={IEEE}
}

@article{HyperspectralNTS,
  author  = {Bajic, Milan and Ivelja, Tamara and Brook, Anna},
  title   = {Developing a Hyperspectral Non-Technical Survey for Minefields via UAV and Helicopter},
  journal = {The Journal of Conventional Weapons Destruction},
  volume  = {21},
  number  = {1},
  year    = {2017},
  article = {11},
  url     = {https://commons.lib.jmu.edu/cisr-journal/vol21/iss1/11}
}

@misc{ENVI,
  title        = {{ENVI}},
  author       = {{L3Harris Geospatial Solutions}},
  year         = {2020},
  note         = {Version Classic 5.7 [Computer software]},
  url          = {https://www.nv5geospatialsoftware.com/Products/ENVI}
}

\end{document}